\documentclass[prd,twocolumn,showpacs,floatfix]{revtex4}
\usepackage{bm}
\usepackage{amssymb}
\begin{document}
\title{Orbital evolution of a particle around a black hole: II.\  
Comparison of contributions of spin-orbit coupling and the self force}
\author{Lior M.\ Burko}
\affiliation{Department of Physics, University of Utah, Salt Lake City,
Utah 84112}
\date{July 31, 2003}
\begin{abstract}
We consider the evolution of the orbit of a spinning compact object in a 
quasi-circular, planar orbit around a Schwarzschild black hole in the
extreme mass ratio limit. We compare the contributions to the orbital
evolution of both spin-orbit coupling and the local self force. Making
assumptions on the behavior of the forces, we suggest that the decay
of the orbit is dominated by radiation reaction, and that the conservative
effect is typically dominated by the spin force. We propose that a
reasonable approximation for the gravitational waveform can be obtained by
ignoring the local self force, for adjusted values of the parameters of
the system. We argue that this approximation will only introduce small
errors in the astronomical determination of these parameters. 
\end{abstract}
 \maketitle

\section{Introduction and Summary}
One of the most promising sources of gravitational radiation that can be
detected by the Laser Interferometer Space Antenna (LISA) \cite{lisa} is
the capture of a compact object by a supermassive black hole. In the
extreme mass-ratio limit this problem can be solved using perturbation
theory, in which the compact object (e.g., a stellar mass black hole
or a neutron star) is treated as a test particle moving on the fixed
background of the central, supermassive black hole.

A test particle, in the absence of self interaction and internal
structure, moves along a geodesic of the spacetime background. When the
particle's mass is much smaller than the typical length scale of the
curvature of the background spacetime, it is useful to use perturbation
theory. The particle's mass (or charge, if any) contributes to the
spacetime curvature, such that the particle moves along a geodesic of a
perturbed spacetime. (These perturbation fields can be interpreted as
those of free gravitational waves, which are produced by the particle at
retarded times. These perturbations are smooth on the particle's worldline
\cite{detweiler-whiting}.) Alternatively, the particle can be construed as
moving along an accelerated, nongeodesic trajectory of the unperturbed 
background spacetime: Let the mass of the companion compact object be
$\mu$, and the mass of the central black hole be $M$, such that $\mu\ll 
M$.  The particle's acceleration at order $\mu$ is
driven by forces at order $\mu^2$ imparted on the particle. To obtain the
corrected orbit of the particle [to $O(\mu)$] one needs to include {\it 
all} the forces, at $O(\mu^2)$, which act on the particle.

Two sources can contribute forces at $O(\mu^2)$: i) the particle's self
force \cite{poisson-review}, and ii) the particle's spin angular
momentum: When the companion of a central black hole is a black hole of
mass $\mu$, the latter carries spin angular momentum $j=s\mu^2$, where
$-1\le s\le 1$. (Neutron stars also have maximal spin angular momentum
which is quadratic in their mass \cite{cook-94}.) The force
imparted on the particle is at order $\mu^2$, which endows the particle
acceleration at order $\mu$. Numerous authors have considered the orbital
evolution under either self force effects \cite{burko-03}
or spin effects \cite{hartl}. However, in order to obtain the particle's
orbit to order $\mu$, one needs to include both the self interaction and
the spin effects. 

The relative importance of the two contributing effects can be evaluated
using an order-of magnitude estimate for $r\gg M$.  
The component of the self-force
responsible for the decay of the orbit (dissipation, ``radiation
reaction'') scales like 
$(\mu/M)^2 (M/r)^5$ \cite{peters-64}. The contribution of the particle's
spin angular momentum to the decay of the orbit can be found from the
Papapetrou equation. Specifically, an order-of magnitude estimate for the
spin force can be found from the fact that it is linear in the particle's
spin, linear in the spacetime curvature, and quadratic in the particle's
four-velocity. (One needs to be careful here, because the various 
components of the four-velocity scale differently in $M/r$. Specifically,
$u^t\sim 1$, whereas $u^{\varphi}\sim (M/r^3)^{1/2}$ --- see below.) 
Taking the typical curvature to scale as $M/r^3$, and the
typical velocity of a particle in a quasi-circular orbit to be given by
Kepler's law, one expects the contribution of the spin force to
dissipation to scale as $(\mu/M)^3 (M/r)^{13/2}$. Consequently, we expect
the decay of the orbit to be dominated by the radiation reaction, with
only small corrections (which we neglect) due to the particle's spin
angular momentum. 

The situation is much different when one considers the
conservative correction to the orbit at $O(\mu)$. The component of the 
self force contributing to the conservative effect scales like $(\mu/M)^2 
(M/r)^6$, whereas the component of the spin force contributing to the
conservative effect scales like $(\mu/M)^2 (M/r)^{7/2}$. (See below the
justification for these scaling laws.) 
Both forces are
at $O(\mu^2)$, but because of the slower drop off with $M/r$, naively one
might expect the spin force under these conditions to be much more
important than the self force. In a strict sense, however, a direct
comparison of the spin force and the self force tells us only little about
the relative importance of the two forces. The reason is that the self
force is gauge-dependent. In particular, one can always choose a gauge in
which the self force vanishes. Consequently, the scaling law used above is
not unambiguous, at least as long as we do not specify the gauge in which
it is written. 

A useful way to compare the spin force and the self force then is to
compare between gauge-{\it independent} quantities. The generated waveform
or the number of cycles that the system spends in a logarithmic interval
of frequency are 
two such quantities. In practice we evolve a quasi-circular, equatorial  
orbit around a Schwarzschild black hole using both forces, or either one,
and compare these gauge-independent quantities. [An initially equatorial
orbit will
remain nearly equatorial because the total angular momentum is conserved,
and the spin angular momentum is much smaller than the orbital angular
momentum. Specifically, the spin angular momentum is at $O(\mu^2)$, such
that its rate of change is at $O(\mu^3)$. Conservation of total angular
momentum imply that changes in the orbital angular momentum are also at
$O(\mu^3)$, whereas the orbital angular momentum itself is at $O(\mu)$. 
Consequently, changes in the direction of the spin vector imply
only very small changes in the direction of the orbital angular momentum,
or very small changes in the orbital plane 
\cite{apostolatos}.] Indeed, we find that our naive
expectations are realized: the conservative effect is controlled by the
spin force, which overwhelms the self force. More general orbits around a
Kerr black hole will include also dissipation due to the spin force and
spin-spin coupling. Here, we ignore such effects, and focus on the
spin-orbit coupling. Our motivation is that for such a simplified orbit we
can already demonstrate our main point: spin forces for astrophysical
systems may be much more important than the self forces, as far as
conservative effects on the orbital evolution are concerned. 

In particular, we find that neglecting the local, conservative self force
in
the construction of templates may introduce only a small error in the
astronomical determination of the parameters of the system. Specifically,
we can match a template made without the local self force almost exactly
to a waveform which does include the local, conservative  self force, but
with a slightly
different value for the spin of the companion. 

In a more general system, the component of the spin force which induces 
the conservative effect also contributes to dissipation. Consider a Kerr
black hole in Boyer-Lindquist
coordinates. The rates of change of the particle's energy, $\dot{E}$, and
angular momentum (in the direction of the spin axis), $\dot{L_z}$, can be
found from the fluxes of energy and angular momentum to infinity and
through the central black hole's event horizon. The rate of
change of Carter's constant, $\dot{Q}$, however, cannot be found using
balance arguments and global conservations laws, because $Q$ is
non-additive. To find $\dot{Q}$ one needs to find first the local forces
which act on the particle. Specifically, 
$$\dot{Q}=G_E({\rm COM},g)\dot{E}
+G_{L_z}({\rm COM},g)\dot{L_z}-2\Sigma\frac{u^r}{u^t}f_r,$$
where $G_E,G_{L_z}$ are certain (known) functions of the constants of
motion (COM) in the absence of dissipation, and the metric $g$
\cite{kennefick-ori}, and $\Sigma=r^2+a^2\cos^2\theta$, $a$ being the
spin parameter of the black hole. To have a
full description of dissipation then, one needs the {\it total} radial
force, $f_r$, which acts on the particle. Similarly to our argument above,
we expect the contribution to $\dot{Q}$ due to the self force to be 
small compared with the contribution of the spin force. 

As generically the companion is expected to spin (and even spin fast), we
propose that a reasonable approximation for the orbital evolution of a
spinning
particle can be found without finding first the local self
force. Specifically,
we propose that a practical way to obtain the orbital evolution (and the
waveforms) to high accuracy could be the following: First, find
$\dot{E},\dot{L_{z}}$ using the fluxes to infinity and down the event
horizon of the central black hole. Next, find $\dot{Q}$ using
$\dot{E},\dot{L_{z}}$ and the {\it spin} contributions to
$f_r$. Undoubtedly, that will introduce an error in the
determination of $\dot{Q}$. However, because of the
smallness of the radial component of the self force with respect to the
radial component of the spin force, we believe that this may be a useful
approximation for the total $\dot{Q}$. To find the orbital evolution one
needs also to include conservative effects \cite{burko-ijmpa}. Again, we
propose to neglect
the contribution of the self force to the conservative effects, and
approximate the full conservative effect by the conservative effect due to
the spin force. 

In the simple problem we consider here the motion is quasi-circular and
equatorial around a Schwarzschild black hole, such that the spin angular
momentum of the companion is aligned (or anti-aligned) with the orbital
angular momentum. It is not hard to let the central black hole be
spinning --- with the spin axis pointing along the same direction as the
companion's spin angular momentum and the orbital angular momentum --- and
include also spin-spin coupling effects. We are hoping to return to
that problem in the future. We note, that when the spins are not aligned
as in our assumptions, the precession of the particle's spin will induce a
changing quadrupole moment, such that the spin force will cause a strong
dissipative effect. The implications of such a spin dissipative effect are 
beyond the scope of the present paper.

Our goal in this paper is more to point out that the accurate
determination of the local self force may perhaps be less crucial for
determination of the orbital evolution of extreme mass-ratio 
binaries than previously thought, than to
give a definite prediction. First, some companions may have only little
spin angular momentum, in which case one would no longer be justified in
neglecting the local self force. Second, and most importantly, our
analysis in what follows is incomplete, in the sense that some of the
necessary pieces of information for carrying out the comparison of the
spin and self
force effects are as yet unknown. We consequently make a number of
assumptions, that allow us to obtain what we believe to be at least a
reasonable order-of-magnitude estimate for the effect of interest. 
We emphasize that our analysis is based on a number of assumptions, which
are as yet unproven. We believe that our main conclusion --- namely, that
spin-orbit coupling may overwhelm the conservative self force effect --- 
is insensitive to the accuracy of these assumptions and is robust. At three 
places our assumptions are speculative (see below for details): First, we
assume that the terms linear in second-order forces in the expression for
the second-order radial velocity are small compared with terms quadratic
in first-order forces. Second, we assume that finite-mass effects in the
luminosity in gravitational waves, which are small in the weak field
regime, are small also in the strong field
regime. Third, we assume that the radial component of the gravitational
self force is proportional to its scalar field counterpart, with a
specific proportionality constant. Of these three assumptions, the last
one is the easiest to test. When this is done, the actual gravitational
self force can be used to replace the assumed expression (\ref{fr-grav}). 
While these assumptions appear to be quite natural to make, they are by no
means guaranteed to be justified under all circumstances. However, even if
any of them turns out to be incorrect, the implications on  our main point
in this paper are not expected to be strong. We therefore believe that our
results are at least reasonable order-of-magnitude estimates, and that
our main point is relevant for realistic extreme mass-ratio binaries. 

The organization of this paper is as follows. In Section \ref{sec2}
we describe the equations of motion and derive the equations for a
quasi-circular, equatorial orbit with aligned spins, using the Papapetrou 
equations and linearizing in the spin covector. Appendix \ref{app-eq} 
includes more details on the definitions of the spin force and the local
self force. In section \ref{sec3} we
derive the perturbative solution for the equations of motion following
the method of Ref.~\cite{burko-03} (paper I), but with slightly different 
definitions and notation, which will make the generalization to a
Kerr black hole simple. Then, we discuss the domain of validity
of the solution. In Section \ref{sec4} we describe the computation of
the evolved orbit and the number of cycles that the system spends
in a logarithmic interval of frequency. Next, in Section \ref{sec5}
we discuss the modeling of the self force. First, we discuss the fitting
of the numerically-derived luminosity in gravitational waves to a smooth
function using a match to two different asymptotic expansions, and discuss
the associated errors, and then we discuss our conjecture for the as-yet
undetermined (radial component of the) self force. Finally, in Section
\ref{sec6}, we compute the orbit and the waveforms, and discuss the
relative importance of the spin force and the self force. 

\section{Equations of motion}\label{sec2} 

\subsection{Equations of motion to order $\mu^2$}
The total force which acts on the particle at order $\mu^2$ is just the
sum of the well known spin force and the self
force. The deviation of the orbit from the geodesic because of the spin
force is at order $\mu$. Its effect on the self force is therefore at
order $\mu^3$, and hence negligible, as we are interested here in the
forces only at order $\mu^2$. Similarly, the deviation of the orbit from
the geodesic because of the self force is also at order $\mu$, such that
its effect on the spin force is again negligible. It is clear then, that
the total force, at order $\mu^2$, which acts on the particle and pushes 
it off the geodesic is just the sum of the self force and the spin
force. Specifically, the equations of motion for a particle of mass
$\mu$, whose center of mass travels along the worldline $z^{\mu}$ with
four-velocity $u^{\mu}=\,dz^{\mu}/\,d\tau$ are given by 
\begin{equation}
\mu\frac{\,Du^{\alpha}}{\,d\tau}=f^{\alpha}\equiv f^{\alpha}_{\rm
SF}+f^{\alpha}_{\rm spin}\, .
\label{eom}
\end{equation}
The expressions we use for $f^{\alpha}_{\rm SF}$ and $f^{\alpha}_{\rm
spin}$ are given in Appendix \ref{app-eq}.

\subsection{Specializing to equatorial motion with an aligned spin}
\subsubsection{General equatorial motion}
The line element, in the usual Schwarzschild coordinates, is given by
\begin{equation}
\,ds^2=-F(r)\,dt^2+\frac{\,dr^2}{F(r)}+r^2\,d\Omega^2
\end{equation}
where $F(r)=1-2M/r$, and 
$\,d\Omega^2=\,d\theta^2+\sin^2\theta\,d\varphi^2$. 
(Notice, that the form of the Papapetrou 
equations of motion depends on the choice of signature.) 

In general, the spin covector precesses along the motion, and the orbit is
not planar. In particular, with arbitrary alignment of the spin covector,
an otherwise equatorial motion will no longer be equatorial. To facilitate
the analysis, let us specialize to a particular solution of the
Papapetrou equations, in which the entire motion is on the equatorial
plane.  
Under the requirement that the entire motion is
equatorial, Eq.\ (\ref{pt}) for the parallel transport of the spin
covector becomes very simple. Specifically, Eq.~(\ref{pt}) becomes 
\begin{equation}
\frac{\,dS^{\alpha}}{\,d\tau}=-\frac{1}{r}u^rS^{\alpha}\delta_{\alpha}^{\theta}
\end{equation}
which implies that $S^{\theta}=s \mu^2/r$, and
$S^t=S^r=S^{\varphi}=0$, $s$ being a constant. Notice that the magnitude
of the spin vector then is $S=s \mu^2$, such that $s$ satisfies $-1<s<1$
for black holes. Equation (\ref{papapetrou}) then becomes
\begin{equation}
\mu\frac{\,Du^t}{\,d\tau}=3\frac{M}{r^3}\frac{r^2}{F(r)}
u^{\varphi}u^rS^{\theta}
\label{eqm_t}
\end{equation}
\begin{eqnarray}
\mu\frac{\,Du^r}{\,d\tau}&=&3\frac{M}{r}F(r)
u^tu^{\varphi}S^{\theta}
\label{eqm_r}
\end{eqnarray}
\begin{equation}
\mu\frac{\,Du^{\varphi}}{\,d\tau}=0
\label{eqm_phi}
\end{equation}
\begin{equation}
\mu\frac{\,Du^{\theta}}{\,d\tau}=0
\label{eqm_theta}
\end{equation}
Notice that the last equation indeed satisfies the requirement for
equatorial motion. The equations of motion
(\ref{eqm_t})--(\ref{eqm_theta}) can be interpreted as an external force,
which is given by their RHS, acting on the particle and pushing it off the
geodesic. This spin force joins the self force, as they are
both at order $\mu^2$. 

\subsubsection{Quasi-circular equatorial orbits}
The equations of motion simplify even further for the case of
quasi-circular orbits. For such orbits the radial motion is perturbatively
small, i.e., the radial velocity is at order $\mu$. That is the case when
an initially circular orbit damps under radiation-reaction
effects. Keeping
terms on the RHS of Eqs.~(\ref{eqm_t})--(\ref{eqm_theta}) to order
$\mu^2$, we find that the only remaining component of the spin force is
the radial component. The spin force we use then is given by
\begin{equation}
f^{\alpha}_{\rm spin}=3\frac{M}{r}\left(1-\frac{2M}{r}\right)
u^tu^{\varphi}S^{\theta}\delta^{\alpha}_{r}+O(\mu^3)\, .
\end{equation}
We thus find that to order $\mu^2$, the only
contribution to the temporal component of the total force comes from the
self force: dissipation is controlled by radiation reaction. The
conservative effects, however, have contributions (via the radial
force) from both the self force and the spin force [as they are both
at $O(\mu^2)$]. In fact, we find that the spin contribution to the radial
force is typically much larger than the self force contribution. We shall
study their relative importance in detail below. 

\begin{figure}
\input epsf
\epsfxsize=8.5cm
\centerline{
\epsfbox{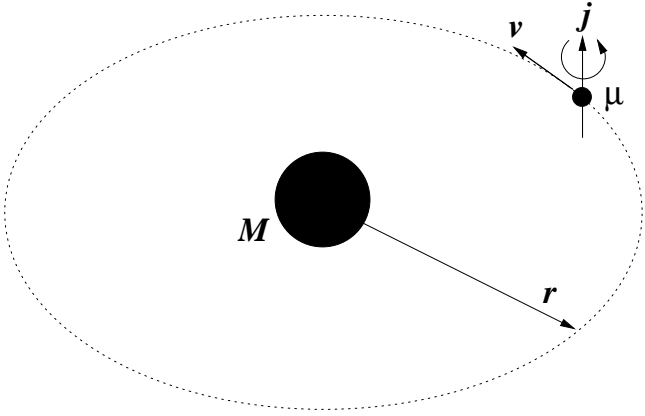}}
\caption{A black hole with mass $\mu$, spin angular momentum ${\bf j}$,
and tangential velocity ${\bf v}$ 
in circular, equatorial motion around a black hole of mass $M$.  
The orbital angular momentum is either aligned
or anti-aligned with the spin angular momentum ${\bf
j}$. In this figures the angular momenta are shown aligned.}
\label{fig1}
\end{figure}

\section{Perturbative solution}\label{sec3}
\subsection{Derivation of the solution}
We next solve the equations of motion (\ref{eom}) perturbatively, for the
case of quasi-circular, equatorial orbit \cite{burko-03}. We use the
normalization
condition for $u^{\alpha}$, namely $u^{\alpha}u_{\alpha}=-1$, to eliminate
$u^t$ from the equations of motion. We next use the $t$ component of the 
equations of motion 
(EOM) to eliminate ${\dot u}^t$. We can simplify the EOM to first-order
(nonlinear) ODEs by taking $\dot{r}=V(r)$, $\dot{x}=Vx'(r)$, and $x$
denotes any quantity. We find the EOM to be
\begin{eqnarray}
VV'&-&3M\frac{V^2}{\Delta}
+\left(M-\omega^2r^3\right)\frac{\Delta}{r^4}\nonumber \\
&=&\frac{1}{\mu {u^t}^2}
\left(\frac{\Delta}{r^2}f_r+r^2\frac{V}{\Delta}f_t\right)
\label{eq1}
\end{eqnarray}
\begin{eqnarray}
V\omega'+2\omega(r-3M)\frac{V}{\Delta}=\frac
{(r-2M)f_{\varphi}+r^3\omega f_t}
{\mu {u^t}^2 \Delta r} 
\label{eq2}
\end{eqnarray}
where $\omega$ is the angular velocity, $\Delta=r^2-2Mr$, 
and  $1/{u^t}^2=1-2M/r-rV^2/\Delta-r^2\omega^2$. 
We next expand Eqs.~(\ref{eq1})--(\ref{eq2}) in powers of 
$\epsilon\equiv\mu/M$: $\omega=\omega_{(0)}+ 
\omega_{(1)}+\omega_{(2)}$, $V= V_{(1)}+ V_{(2)}$, and $a_{i}=
a^{(1)}_i+ a^{(2)}_i$. Here, $x_{(j)}$ denotes the term in $x$ which is
at $O(\epsilon^{j})$, and $a_i$ being the self acceleration. We then
expand the self force as $f_{i}^{\rm SF}= f^{(1)}_i+ f^{(2)}_i$, where
$f^{(j)}_i=\mu a^{(j)}_i$. We then expand Eqs.~(\ref{eq1})--(\ref{eq2}),
and solve perturbatively order by order. 

The zeroth order term of
Eq.~(\ref{eq1}) recovers Kepler's law. Specifically, it yields 
\begin{equation}
\omega_{(0)}=\frac{M^{\frac{1}{2}}}{r^{\frac{3}{2}}}\, .
\end{equation}
The first order corrections to Kepler's law are obtained from the terms at
$O(\epsilon)$ of Eq.~(\ref{eq1}). Specifically, we find that 
\begin{eqnarray}
\omega^{(1)}=\frac{rf_r^{(1)}}{2\mu}\frac{r^3\omega_{(0)}^2
-(r-2M)}{r^3\omega_{(0)}}\, .
\end{eqnarray}
We next define  
\begin{eqnarray}
\varpi^2 :=  r^2\Delta\omega'_{(0)}+2r^2(r-3M)\omega_{(0)}\, .
\end{eqnarray}
The first order term of the radial velocity is obtained from the terms at
$O(\epsilon)$ of Eq.~(\ref{eq2}):
\begin{eqnarray}
V^{(1)}=&-&\frac{1}{\mu\varpi^2}\left[r^3\omega_{(0)}^2-
(r-2M)\right]\nonumber \\ &\times&
\left[(r-2M)f_{\varphi}^{(1)}+r^3\omega_{(0)}f_t^{(1)}\right]\, .
\end{eqnarray}
The second order correction to the radial velocity is found from the 
terms at $O(\epsilon^2)$ of Eq.~(\ref{eq2}):
\begin{eqnarray}
V^{(2)}&=&-\frac{1}{\mu\varpi^2}\left\{r^3\omega^{(1)}f_t^{(1)}
\left[3r^3\omega_{(0)}^2-(r-2M)\right]\right.
\nonumber \\
&+& \left. 2r^3(r-2M)\omega^{(1)}f_{\varphi}^{(1)}
\omega_{(0)}\right.\nonumber \\
&+&\left. \mu r^2 V^{(1)} \left[
2\omega^{(1)}\left(r-3M\right)+\Delta\omega'_{(1)}\right] 
\right. \nonumber \\
&+&\left. r^3\omega_{(0)}\left[r^3\omega_{(0)}^2-(r-2M)\right]f_t^{(2)}
\right.\nonumber \\
&+&\left.
(r-2M)f_{\varphi}^{(2)}\left[r^3\omega_{(0)}^2-
(r-2M)\right]\right\}\label{v2}
\end{eqnarray}
As we shall see below, $\omega^{(2)}$ does not contribute at the order to
which we are solving the EOM. The explicit perturbative solution of the EOM 
is listed in Appendix B. 

We remark that we express all quantities here as functions of $r$. For
other purposes, e.g., for analysis of detected signals, it is frequently
more convenient to express quantities as functions of $\omega$. It is easy
to translate our expressions to functions of $\omega$ by noting that 
$\,d\omega=\omega '\,dr$.

\subsection{Approximating the solution and its validity regime}
Notice that each term inside the curly brackets in Eq.~(\ref{v2}) is at
$O(\mu^3)$. These terms come in two kinds: first, there are terms which
are quadratic in $f_{\mu}^{(1)}$ (or their gradients), and second, there
are terms which are linear in $f_{\mu}^{(2)}$. The latter, of course are
as yet unknown. Their derivation requires second-order perturbation
theory, and the extension of regularization techniques to that order. One
should therefore view Eq.~(\ref{v2}) as a formal expression. However, we
propose that the expression for $V^{(2)}$ is dominated by the terms
quadratic in $f_{\mu}^{(1)}$, such that neglecting the terms involving 
$f_{\mu}^{(2)}$ introduces only a small error: Consider the $t,r$
components of the four-acceleration and the four-force. Expand the
four-acceleration as $a=a^{(1)}+a^{(2)}+...$ where $a^{(n)}$ is at order 
$\epsilon^n$. Then obviously $a^{(2)}\propto {a^{(1)}}^2$,
with some proportionality constant (with dimensions of $1/M$) 
which we expect to be neither very large nor very small. (Recall
that we solve perturbatively, such that $a^{(1)}\sim \mu/M^2$. In this
case $a^{(2)}\ll a^{(1)}$.) The force $f_{\nu}^{(2)}=\mu
a_{\nu}^{(2)}$  can be written as
$f_{\nu}^{(2)}\propto\mu {a_{\nu}^{(1)}}^2$. Substituting
$a_{\nu}^{(1)}=f_{\nu}^{(1)}/\mu$ we then find that
$f_{\nu}^{(2)}\propto\mu^{-1} {f_{\nu}^{(1)}}^2$. As $f_{\nu}^{(2)}$ has
the same dimensions as $f_{\nu}^{(1)}$ (namely, it is dimensionless in a
normalized basis), we expect that $f_{\nu}^{(2)}=\alpha_{\nu} 
M\mu^{-1} {f_{\nu}^{(1)}}^2$, (no summation over repeated indices
implied) where 
$\alpha_{\nu}$ is an (unknown) dimensionless function of $r/M$. Next
consider the $\varphi$ component. We 
expect $f_{\varphi}^{(1)}= -f_{t}^{(1)}/\omega$. Substituting Kepler's law
for $\omega$, and repeating our arguments above, we find that 
$f_{\varphi}^{(2)}=\alpha_{\varphi}(M/r)^{3/2}\mu^{-1}
{f_{\varphi}^{(1)}}^2$, where again $\alpha_{\varphi}$ is a dimensionless
function of $r/M$. 

Consider now the terms in Eq.~(\ref{v2}). The coefficients
of the terms proportional to $f_{\mu}^{(2)}$ are typically much smaller
than the coefficients of the terms quadratic in $f_{\mu}^{(1)}$. For
example, compare the
term involving $f_t^{(2)}$ and the 
term involving  $f_t^{(1)}f_r^{(1)}$ in Eq.~(\ref{v2-1}). The ratio of
these two terms is
\begin{eqnarray}
{\cal R}:= \frac{B_tf_t^{(2)}}{A_{tr}f_t^{(1)}f_r^{(1)}}&\sim&
\alpha_t\frac{(M/\mu)B_t{f_t^{(1)}}^2}{A_{tr}f_t^{(1)}f_r^{(1)}}\nonumber
\\
&\sim&\alpha_t\frac{(M/\mu)B_tf_t^{(1)}}{A_{tr}f_r^{(1)}}\, .
\end{eqnarray}
Next, for quasi-circular orbits at $r\gg M$, $f_t^{(1)}/f_r^{(1)}\sim  
(M/r)^{1/2}$, such that 
\begin{eqnarray}
{\cal R}
&\sim&\alpha_t\frac{(M/\mu)B_t}{A_{tr}}\left(\frac{M}{r}\right)^{1/2}
\nonumber \\
&\sim& \frac{\alpha_t}{2}\frac{r-6M}{r-3M}
\left(\frac{M}{r}\right)^{5/2}\, .
\end{eqnarray}
Here, $B_t=2r(r-3M)/[\mu(r-6M)]$ and $A_{tr}=4r^3(r-3M)^2/[\mu^2M(r-6M)^2]$ 
are the coefficients of the terms proportional to $f_t^{(2)}$ and
$f_t^{(1)}f_r^{(1)}$, respectively, in $V^{(2)}$. 
Demanding that ${\cal R}\ll 1$ introduces then retrictions on the function 
$\alpha_t$. Specifically, for $r\gg M$ our assumption is justified if
$\alpha_t$ increases as a function of $r/M$ slower than
$(r/M)^{5/2}$. (Even if $\alpha_t$ violates this condition, our assumption
can still be valid in a neighborhood of a point $r$ at which it is
valid. If the condition is satisfied, our assumptions are valid globally.)  
Presently, as was already
discussed above, we have no knowledge of the functions $\alpha_{\mu}$. 
We proceed by assuming that
$\alpha_{\mu}$ are such that indeed the terms proportional to
$f_{\mu}^{(2)}$ are small compared with the terms quadratic in
$f_{\mu}^{(1)}$ in Eq.(\ref{v2}). 
Similarly, we can introduce analogous constraints on
$\alpha_r$ and $\alpha_{\varphi}$ such that 
any term linear in $f_{\mu}^{(2)}$ in Eq.~(\ref{v2}) would indeed be 
much smaller than any term quadratic in $f_{\mu}^{(1)}$. 
Although our discussion above is limited to $r\gg M$, we assume the
applicability of our conclusions for all $r>6M$. We therefore
neglect all the terms in Eq.~(\ref{v2}) which are linear in
$f_{\mu}^{(2)}$. 

We emphasize that even if our assumption that all the terms in
Eq.~(\ref{v2}) which are linear in $f_{\mu}^{(2)}$ are negligible compared
with terms which are quadratic in $f_{\mu}^{(1)}$ turns out to be
incorrect, our main conclusion in this paper is still likely to be
relevant for at least a portion of the orbit (see below).

The solution we find for the EOM is a perturbative solution about a
circular orbit. So long as the orbit does not get too far from circularity
we therefore expect our solution to be valid. However, when the particle
arrives at the innermost stable circular orbit (ISCO) at $r=6M$ the orbit
changes from an adiabatic, quasi-circular orbit
to a plunge. (We do not consider here the change in the ISCO itself under
radiation reaction and spin-orbit effects.) 
When that happens the radial velocity can no longer be
considered as small, and the perturbative approach to the solution (i.e.,
expanding the solution about a circular orbit) breaks down. To find where
our perturbative approach loses its validity let us compare the magnitude
of $V^{(2)}$ with the magnitude of $V^{(1)}$. We consider the perturbative
approach as valid only if $V^{(2)}$ is smaller than
$V^{(1)}$. Specifically, let us introduce the condition that 
$|V^{(2)}/V^{(1)}|\lesssim\gamma$ for the perturbative approach to be
valid, where $\gamma$ is a positive constant smaller than unity. (In
practice, we arbitrarily fix $\gamma=0.2$.) For a typical term in
$V^{(2)}$ we take the term proportional to $f_t^{(1)}f_r^{(1)}$ in
Eq.~(\ref{v2-1}), and for a
typical term in $V^{(1)}$ we take the term proportional to $f_t^{(1)}$ in
Eq.~(\ref{v1-1}). Their ratio is
\begin{equation}
\left|\frac{V^{(2)}}{V^{(1)}}\right|\approx \frac{2}{\mu M}
\frac{r^2(r-3M)}{r-6M}\left|f_r^{(1)}\right|\lesssim\gamma\, .
\end{equation}
As $f_r^{(1)}$ is bounded, clearly this inequality is violated at some
value of $r>6M$. However, as $f_r^{(1)}$ is at $O(\mu^2)$, the domain of
validity of the perturbative approach extends to smaller values of $r$ the
smaller $\mu$.

\section{Computation of the orbit and the waveform}\label{sec4}
In the previous section we found $V(r)$ to order $\mu^2$ and $\omega(r)$
to order $\mu$ for quasi-circular equatorial orbits. Next, we find the
orbit by computing
\begin{equation}
t(r)=\int_{r_{\rm initial}}^{r}\frac{\,d{\tilde r}}{V({\tilde r})}
\;\;\;\;{\rm and}\;\;\;\;
\varphi(r)=\int_{r_{\rm initial}}^{r}\frac{\omega({\tilde r})\,d{\tilde
r}}{V({\tilde r})}\, ,
\end{equation}
where the integrands are evaluated using the perturbative solution for
$V(r)$ and $\omega(r)$. In practice, we solve these integrals using a
fourth-order Runge-Kutta integrator. This yields the triad
$r,t(r),\varphi(r)$, which we invert to $t,r(t),\varphi(t)$, which is just
the required orbit. 

Next, we compute the number of cycles ${\cal N}_{\rm cyc}$ spent in a
logarithmic interval of frequency $f$, 
$\,d{\cal N}_{\rm cyc}/\,d{\rm ln}f$, by noting that 
$\,d{\cal N}_{\rm cyc}/\,d{\rm
ln}f=\omega^2/(\pi{\dot\omega})$. Re-expressing the last equation as 
$\,d{\cal N}_{\rm cyc}/\,d{\rm  
ln}f=\omega^2/(\pi\omega' V)$ and using the perturbative solution for
$V(r)$ and for $\omega(r)$, we find that 
\begin{eqnarray}
\frac{\,d{\cal N}_{\rm cyc}}{\,d{\rm ln}f}=
-\frac{2}{3\pi}\left(\frac{M}{r}\right)^{\frac{1}{2}}\frac{1}{V^{(1)}}
\left\{1+ \left(\frac{r}{M}\right)^{\frac{1}{2}}r^3 \right.
\nonumber \\
\times\left.
\left[2\frac{\omega^{(1)}}{r^2}
+\frac{2}{3}
\frac{\omega^{'(1)}}{r}\right]-\frac{V^{(2)}}{V^{(1)}}+O(\mu^2)
\right\}
\label{d_N}
\end{eqnarray}

Finally, the total change in the number of cycles can be found by
integrating
\begin{equation}
\Delta{\cal N}_{\rm cyc} := \int_{r_{\rm initial}}^{r}
\frac{\omega '}{\omega}\frac{\,d{\cal N}_{\rm cyc}}{\,d{\rm ln}f}\,d
{\tilde r}\, .
\label{Delta_N}
\end{equation}

\section{Modeling the self force}\label{sec5}
\subsection{Fitting the luminosity in gravitational waves to a smooth
function}
The {\it local} gravitational self force has not been computed yet for a
point particle in motion around a Schwarzschild black hole, not even for
quasi-circular, equatorial orbits. (An exception is a radial plunge, for
which it has been calculated \cite{barack-lousto}.) However, the radiation
reaction for quasi-circular, equatorial orbits in Schwarzschild (and also
for other types of orbits, including in Kerr) was calculated using balance
arguments, and the results for the rate of change of the constants of
motion can be translated into the local self force. As the local self
force is not available to us, we shall proceed by using results obtained
from the non-local approach. That approach can give us the dissipative
part of the self force, but not the conservative part, which is not
encoded in the gravitational-wave luminosity. We will therefore estimate
the conservative piece of the self force by analogy the the case of scalar
field self interaction. Although we do not expect our estimate to be
accurate, we nevertheless hope that it can still provide us with a crude
order-of-magnitude estimate for the actual conservative piece of the self
force. 

In practice, we take the results for the luminosity of gravitational waves
(i.e., the flux to infinity) for a test mass in circular orbit around a
Schwarzschild black hole \cite{cutler}.  These results are given in the
form of a table,
which gives $\,dE/\,dt(r)$. For the purpose of integrating a
slowly-evolving orbit, we need a smooth function. We can obtain such a
function by fitting the data in the table to a smooth function. The most
natural functional form to use is that of the post-Newtonian
(PN) expansion. While the PN expansion converges very rapidly at large
distances, it does so only very poorly close to the ISCO, where the motion
is very relativistic. On the other hand, a general polynomial fit
converges very rapidly at small distances, but poorly at large
distances. We find that the two asymptotic expansions match at an overlap
region, such that a matched asymptotic expansion method can be very
useful. 

In practice, we use the 3.5PN expansion for a test particle at
large distances. The flux of energy to infinity is given by
\cite{tagoshi,blanchet}
\begin{eqnarray}
\frac{\,dE}{\,dt}^{3.5{\rm PN}}&=&\frac{32}{5}\frac{\mu^2
M^3}{r^5}\left\{1-
\frac{1247}{336}v^2+4\pi v^3-\frac{44711}{9072}v^4
\right. \nonumber \\
&-&\frac{8191}{672}\pi v^5+\left[\frac{6643739519}{69854400}-
\frac{1712}{105}\gamma_{\rm e}+\frac{16}{3}\pi^2\right.\nonumber \\
&-&\left.\left.\frac{856}{105}\ln(16 v^2)\right]v^6-\frac{16285}{504}\pi
v^7\right\}\, ,
\end{eqnarray}
where $v=(M\omega)^{1/3}$ is the orbital velocity, and where $\gamma_{\rm 
e}$ is Euler's constant. 

At small distances we fit the luminosity to a polynomial of the form 
\begin{equation}
\frac{\,dE}{\,dt}^{\rm PL}=
\frac{32}{5}\frac{\mu^2M^3}{r^5}\sum_{k=0}^{n}C_kr^k
\label{fit-pol}
\end{equation}
with coefficients $C_k$ given in Appendix \ref{app2}.

Figure \ref{fig2} shows the temporal component of the self force and the
relative errors in its determination in terms of both methods. As
expected, the relative error in the 3.5PN expansion is very small at large
distances, but grows rapidly at small distances. Similarly, the power-law
approximation yields highly accurate results at small distances, but is
inaccurate at large distances. There is an overlap region, for 
$12\lesssim r/M \lesssim 24$, where the two expansions have comparable
accuracy. Specifically, the 3.5PN expansion yields at $r=12M$ a relative
error of $1-(\,dE/\,dt)^{3.5{\rm PN}}/(\,dE/\,dt)\approx 4\times 10^{-4}$,
and the power-law expansion yields at $r=24M$ a relative error of 
$1-(\,dE/\,dt)^{\rm PL}/(\,dE/\,dt)\approx 1.1\times 10^{-3}$. Allowing
our accuracy in the determination of the luminosity to be $1$ part in
$10^{3}$, we evaluate the self force at large distances ($r/M >
24$) using the 3.5PN expansion, and at small distances ($r/M < 12$) using
the power-law expansion. For the region $12\leqslant r/M \leqslant 24$ we
use a sliding average of the two methods, i.e., we simply take
\begin{equation}
\frac{\,dE}{\,dt}^{\rm mix}=\frac{\,dE}{\,dt}^{\rm PL}\frac{24M-r}{12M}
+\frac{\,dE}{\,dt}^{3.5{\rm PN}}\frac{r-12M}{12M}\, .
\end{equation}
This guarantees that our fit to the luminosity, and to the temporal
component of the self force, will nowhere be worse than our tolerance.

\begin{figure}   
\input epsf
\epsfxsize=8.5cm
\centerline{
\epsfbox{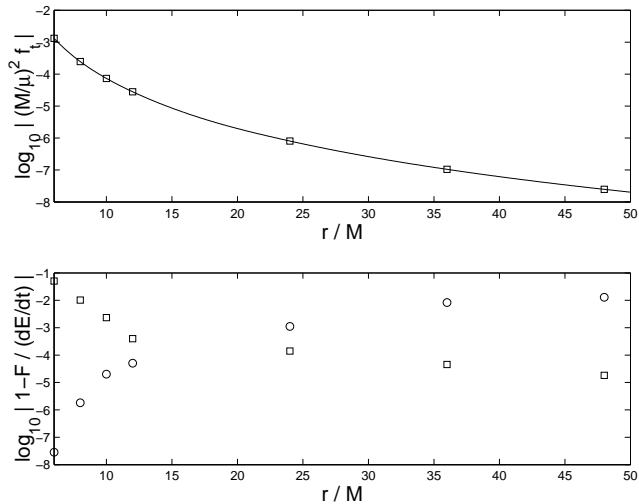}}
\caption{Luminosity in gravitational waves. Upper panel: The temporal
component of the self force [Eq.~(\ref{ft})] as a function of $r/M$. The
numerical data are represented by $\square$'s, and the fitted curve by the
solid line. Lower panel: The relative error in the determination of the
luminosity $F$ using the 3.5PN
expansion [$\square$, $ F := (\,dE/\,dt)^{3.5{\rm PN}}$] and the power-law
expansion [$\circ$, $F := (\,dE/\,dt)^{\rm PL}$]. }
\label{fig2} 
\end{figure}

\subsection{Error analysis}
As we already discussed in the preceding subsection, the error in the
determination of the luminosity in gravitational waves for a test mass is
no worse that $10^{-3}$. The other sources of errors which affect our
determination of the $t$-component of the self force are finite mass
effects (i.e., the effect of a finite $\mu/M$), and the flux of energy
through the black hole's event horizon. 

The total rate of change of energy of the particle is the sum of the
luminosity in gravitational waves (i.e, the flux of energy to
infinity) and the flux in energy through the event horizon of the black
hole. Specifically, 
\begin{equation}
\frac{\,dE}{\,dt}^{\rm TOT}=\frac{\,dE}{\,dt}^{\infty}+
\frac{\,dE}{\,dt}^{\rm EH}\, .
\end{equation}
As $(\,dE/\,dt)^{\rm EH}\approx \alpha v^8(\,dE/\,dt)^{\infty}$, with
$\alpha$ being a constant of order unity \cite{poisson-sasaki}, the flux
of energy though the
event horizon where it is the greatest (when the particle is at $r=6M$) is
smaller than the
flux to infinity by a factor of $8\times 10^{-4}$. Consequently,
neglecting the absorption by the black hole introduces an error in the
determination of the $t$-component of the self force which is compatible
with our tolerance.  

The other source for errors in the determination of the $t$-component of
the self force are finite-mass effects. The finite-mass effects are easy
to estimate for large distances, where they are fully known at the 3.5PN
level. (Recall however the ambiguity in determination of the finite-mass
effect of the 3PN terms.) The error introduced by neglecting the
finite-mass effects in the 3.5PN expansion is $({\rm a\;\;few})\times
\mu/M$. The error will be compatible with our tolerance if we take the
mass ratio to be $\mu/M\lesssim ({\rm a\;\;few})\times 10^{-4}$. We shall
therefore restrict the systems we study here to compatible mass ratios. We
cannot estimate finite-mass effects close to the ISCO, as the luminosities
available to us were obtained from a linearized analysis. We are
encouraged, however, that at all PN orders (up to 3.5PN) the errors
introduced by neglecting finite-mass effects are comparable. If that
behavior persists at all PN orders, we are guaranteed that our
approximation is valid also near the ISCO. Although we do not know whether
this is indeed the case, we hope that our approximation remains
qualitatively valid also near the ISCO, and that it does not introduce
errors significantly bigger than our tolerance.

\subsection{Determination of the self force}
The $t$-component of the self force is easily determined by
\begin{equation}\label{ft}
f_t^{\rm SF}=\frac{\,dt}{\,d\tau}\; \frac{\,dE}{\,dt}\, .
\label{ft-flux}
\end{equation}
In Eq.~(\ref{ft-flux}) $\,dE/\,dt$ is the flux of energy to infinity. The
rate of change of the particle's energy then is $-\,dE/\,dt$. 
Notice, that we only need to determine $f_t^{(1)}$, such that in
Eq.~(\ref{ft}) the factor $\,dt/\,d\tau$ can be evaluated for geodesic
motion, i.e., $\,dt/\,d\tau=1/(1-3M/r)^{1/2}$. The $\varphi$-component of
the self force can be determined from its proportionality to the
$t$-component.

The $r$-component of the self force cannot be found by using non-local
methods. Instead, it needs to be evaluated using a fully local
calculation. As the results of such a calculation are as yet unavailable
to us, we shall instead use a crude order-of-magnitude estimate for
$f_r^{\rm SF}$. Because of the smallness of the effects of this component,
and because it is {\it a posteriori} found to be much smaller than the
magnitude of the $r$-component of the spin force, we do not expect our
crude estimate to be problematic. Note, that once $f_r^{\rm SF}$ is
calculated, it can easily be used to replace our estimate
here. Specifically, we
estimate the radial component of the self force by using the known results
for the radial self force on a scalar charge $q$ in circular orbit around
a Schwarzschild black hole \cite{burko-00,detweiler}. In order to do that
estimate, recall that for $r\gg M$, 
${_{\rm grav}f}_t^{\rm SF}/{_{\rm scalar}f}_t^{\rm SF}\approx 
\frac{96}{5}(\mu/q)^2(M/r)$ \cite{peters-64,burko-00}. We next assume that
an analogous scaling is satisfied also by $f_{\mu}^{(2)}$ (which is
entirely conjectural, although not implausible). Consequently, because 
$f_{\alpha}^{\rm SF}u^{\alpha}=0$, it is reasonable that the same
proportionality is satisfied also by the $r$-components,
$f_{r}^{(1)}$. For strict
circular orbits the radial velocity vanishes, such that no restrictions
on $f_{r}^{(1)}$ can be 
applied, but for quasi-circular orbits the orthogonality of the force and
the 
orbit imply that it is not implausible that such a proportionality is 
satisfied. In view of this argument, we assume 
\begin{equation}
{_{\rm grav}f}_r^{\rm SF}\approx\beta\frac{\mu^2}{q^2}\frac{M}{r}
\times {_{\rm scalar}f}_r^{\rm SF}\, .
\label{fr-grav}
\end{equation}
Although we expect $\beta=96/5$, in practice we parametrize our results
with $\beta$. Although our arguments may apply to the far field region
only, we extrapolate the scaling relation (\ref{fr-grav}) for all values of
$r$. This choice, in the absence of numerical data, appears to us to be
reasonable, at least as an approximation for the actual radial component
of the self force. 

\section{Calculated orbits and waveforms}\label{sec6}

\begin{figure}
\input epsf
\epsfxsize=8.5cm
\centerline{  
\epsfbox{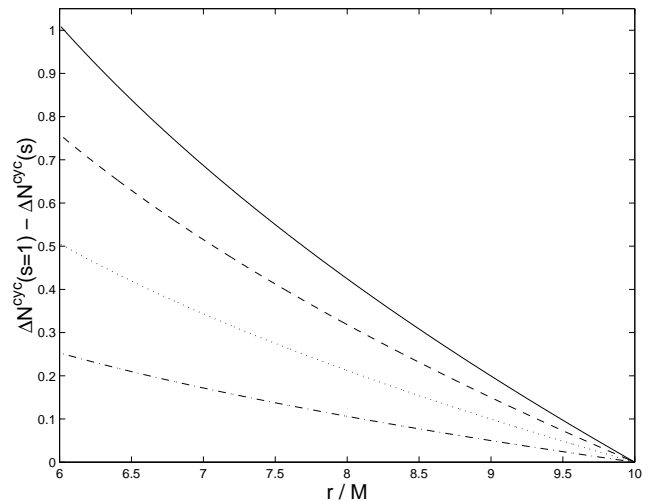}}
\caption{The
difference between $\Delta {\cal N}_{\rm cyc}$ which was obtained for
$s=1$ and a number of values of $s$, as a function of $r/M$. The data are 
for a system with $\mu=5\times
10^{-4}M$ that starts decaying at $r=10M$. Shown are the results for
$s=-1$ (solid curve), $s=-0.5$ (dashed curve), $s=0$ (dotted curve), 
and $s=0.5$ (dashed-dotted curve).} 
\label{dn-spin}
\end{figure}

In order to compare between different evolutions we use $\,d{\cal N}_{\rm
cyc}/\,d\ln f$, the number of cycles ${\cal N}_{\rm cyc}$ that the system
spends in a logarithmic interval of frequency $f$, as given in
Eq.~(\ref{d_N}). The total number of cycles that the system undergoes
between two values of $r$, $\Delta {\cal N}_{\rm cyc}$ (this, of course,
can be translated to the number
of cycles between two values of the frequency $f$), is found by using
Eq.~(\ref{Delta_N}). First, we discuss the contribution to $\Delta {\cal
N}_{\rm cyc}$ from the spin-orbit coupling. Figure \ref{dn-spin} shows the
difference between $\Delta {\cal N}_{\rm cyc}$ which was obtained for $s=1$
and a number of values of $s$, for a system with $\mu=5\times 10^{-4}M$ 
that starts decaying at $r=10M$. The maximum effect is obtained when
$s=-1$, where it is just above $1$ cycle at $r=6M$. This effect is at
$O(\mu^0)$, i.e., it is independent of the mass ratio \cite{burko-03}. 
That is, ignoring the spin force would result in a maximal error of about
a full cycle, which could reduce the correlation integral of the signal
with a theoretical template considerably. (Notice that the total number of
cycles that the system undergoes is ${\cal N}_{\rm cyc}\approx 2.6\times
10^3$, and that ${\cal N}_{\rm cyc}\propto\mu^{-1}$.) 

\begin{figure}  \label{wf}
\input epsf
\epsfxsize=8.5cm
\centerline{
\epsfbox{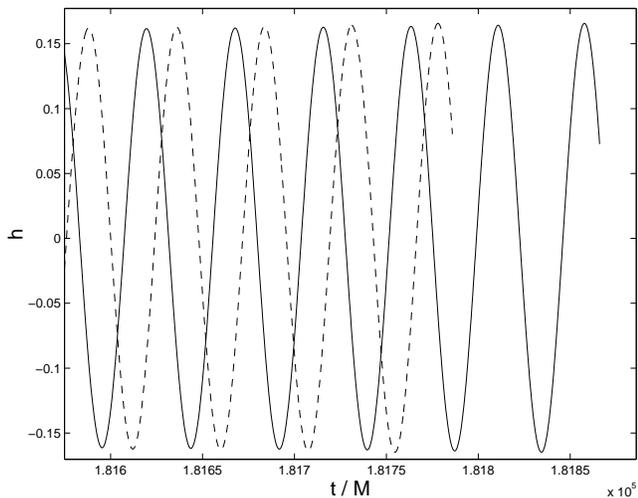}}
\caption{The waveform -- $h$ as a function of $t/M$. The data are shown for
$\mu=5\times 10^{-4}M$. We show the last few oscillations before
$r=6M$. Plotted are the waveforms for two values of $s$: $s=1$
(solid curve) and $s=-1$ (dashed curve). The two curves were in phase at
$t=0$ ($r=10M$).} 
\end{figure}

The waveforms, which we present using the ``restricted waveform''
approximation \cite{damour}, are displays in Fig.~\ref{wf}. (Our
calculation of the $\Delta {\cal N}_{\rm cyc}$ is independent of the
assumption of the ``restricted waveform'' approximation.)  Figure 
\ref{wf} indeed shows that the maximal ambiguity in phase due to ignoring
the spin-orbit coupling is about two wavelengths. 

\begin{figure}
\input epsf
\epsfxsize=8.5cm
\centerline{   
\epsfbox{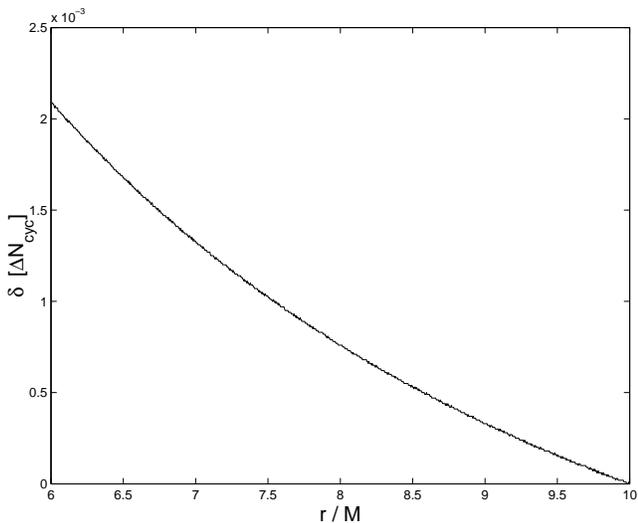}}
\caption{$\delta [\Delta {\cal N}_{\rm cyc}]$ --- the difference between
$\Delta {\cal N}_{\rm cyc}$ with the
inclusion of the self force, and $\Delta {\cal N}_{\rm cyc}$ without its
inclusion --- as a function of $r/M$. The data shown are for $\mu=5\times  
10^{-4}M$, $s=1$, and the system starts decaying at $r=10M$.}
\label{dn-sf}
\end{figure}

\begin{figure}
\input epsf
\epsfxsize=8.5cm
\centerline{
\epsfbox{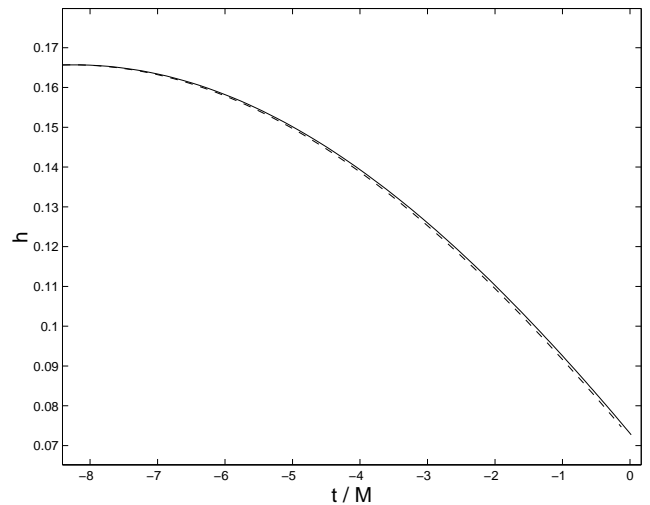}}
\caption{The waveforms for $\mu=5\times
10^{-4}M$ and $s=1$. Solid curve: including the effects of the self
force. Dashed curve: neglecting the effect of the self force. Here, we used
$\beta=1$, and the two waveforms were in phase at $r=10$. Shown is a
fraction of the last oscillation before $r=6M$. Wr set $t=0$ at $r=6M$.}
\label{wwofr}
\end{figure}

In order to examine how important that self force may be we fix $s=1$, and
compute $\delta [\Delta {\cal N}_{\rm cyc}]$, the difference in the total
number of cycles for the case in which
we include the self force in the calculations of the orbital evolution, and
the case in which we do not. The results, for $\beta=1$, are presented in
Fig.~\ref{dn-sf}. We find that the difference in the total number of cycles
is $2\times 10^{-3}\beta$ at $r=6M$.  Even for $\beta\approx 20$ the 
difference in the total number of cycles is only $4\times 10^{-2}$, which
is very small indeed. This effect is at $O(\mu^0)$ \cite{burko-03}. 
Figure \ref{wwofr} displays a fraction of the
last oscillation in the waveforms with and without the inclusion of
the self force (for $s=1$) before $r=6M$. 

\begin{figure}
\input epsf
\epsfxsize=8.5cm
\centerline{
\epsfbox{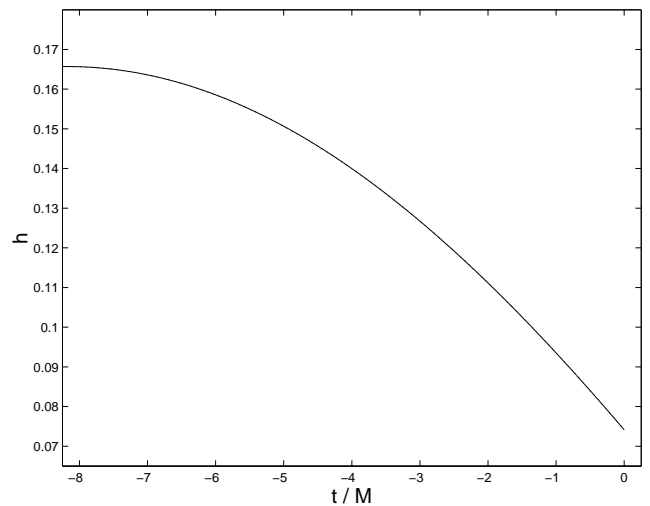}}
\caption{The waveforms for $\mu=5\times
10^{-4}M$. Solid curve: excluding the effects of the self
force for $s=1$. Dashed curve: including the effect of the self
force for $s=0.997$. Here, we used
$\beta=1$, and the two waveforms were in phase at $r=10$. Shown is a
fraction of the last oscillation before $r=6M$. The two waveforms are
indistinguishable on this scale.} \label{cor}
\end{figure}

Most importantly, we can make the waveform without the inclusion of the
self force coincide almost exactly with the waveform with its inclusion, if
we modify the value of $s$ by a small number which is comparable to the 
difference in the number of cycles for the cases with and without the
inclusion of the self force. This situation is shown in Fig.~\ref{cor},
which displays two waveforms: one waveform is the same as the waveform
without including the self force effect in Fig.~\ref{wwofr}. The other
waveform is a waveform with the self force effect, but with a corrected
value of the spin $s$. The two waveforms overlap almost exactly over the
entire wave train, and in particular during the last fraction of the orbit
before $r=6M$. 

Therefore, we propose that neglecting the self
force will only introduce a small error in the determination of the spin
rate of the companion, at the order of $\delta [\Delta {\cal N}_{\rm
cyc}]$. 

\section*{Acknowledgments}
I am indebted to Richard Price for many invaluable discussions. This
research was supported by the National Science Foundation through grant
No.~PHY-9734871.

\begin{appendix}

\section{Expressions for $f^{\alpha}_{\rm SF}$ and
$f^{\alpha}_{\rm spin}$}\label{app-eq}
In this Appendix we write explicitly the expressions for $f^{\alpha}_{\rm
SF}$ and for $f^{\alpha}_{\rm spin}$ which appear in Eq.~(\ref{eom}). 
Specifically,  
\begin{eqnarray}
f^{\alpha}_{\rm SF}&=&\mu^2k^{\alpha\beta\gamma\delta}   
\lim_{\tau\to 0^-}\int_{-\infty}^{\tau}
\nabla_{\delta}G_{\beta\gamma\beta '\gamma '}[z^{\mu}(\tau);
z^{\mu}(\tau ')]\nonumber \\
&\times &u^{\beta '}(\tau ')u^{\gamma '}(\tau ')\,d\tau'
\end{eqnarray}
where
\begin{eqnarray}
k^{\alpha\beta\gamma\delta}&=&\frac{1}{2}g^{\alpha\delta}u^{\beta}u^{\gamma}
-g^{\alpha\beta}u^{\gamma}u^{\delta}-\frac{1}{2}u^{\alpha}u^{\beta}  
u^{\gamma}u^{\delta}
\nonumber \\
&+&\frac{1}{4}g^{\beta\gamma}u^{\alpha}u^{\delta}+
\frac{1}{4}g^{\alpha\delta}g^{\beta\gamma}
\end{eqnarray}
and $G_{\beta\gamma\beta '\gamma '}[z^{\mu}(\tau);z^{\mu}(\tau ')]$ is
the two-point retarded Green's function \cite{poisson-review}.

The spin force is given  (at the pole-dipole approximation, i.e., taking
into consideration only the mass monopole and the spin dipole, neglecting
higher multipoles, such as the tidal coupling, which is a mass quadrupole
effect) by the
Papapetrou equations
\begin{equation}\label{p1}
\frac{\,Dp^{\alpha}}{\,d\tau}=-\frac{1}{2}S^{\sigma\mu}u^{\nu}
R^{\alpha}_{\;\;\nu\sigma\mu}
\end{equation}
and
\begin{equation}\label{p2}  
\frac{\,DS^{\alpha\beta}}{\,d\tau}=2u_{\rho}u^{[\alpha} 
\frac{\,DS^{\beta ]\rho}}{\,d\tau}
\end{equation}
where $p^{\mu}=\mu u^{\alpha}-u_{\beta}\frac{\,DS^{\alpha\beta}}{\,d\tau}$
is the particle's four-momentum, $S^{\alpha\beta}$ is the skew-symmetric
spin tensor of the particle, which is given by
\begin{equation}
S^{\alpha\beta}=2\int_{\Sigma}(x^{[\alpha}-z^{[\alpha})T^{\beta ]\gamma}
\,d\Sigma_{\gamma}\, ,
\end{equation}
where $\Sigma$ is an arbitrary spacelike hypersurface. Notice that the RHS
of Eq.~(\ref{p2}) is just $2p^{[\alpha}u^{\beta ]}$.

As the mass of the particle $\mu\ll M$, and as the spin of the particle is
consequently also small, we shall approximate the Papapetrou equations by
considering the spin force to leading order, that is to order $\mu^2$
\cite{apostolatos}.
We next introduce the Mathisson-Pirani spin supplementary condition (which
physically identifies the particle's center of
mass) $S^{\alpha\beta}u_{\beta}=0$ and the Pauli-Lubanski spin covector
$S_{\alpha}$, defined by
\begin{equation}
S_{\alpha}=\frac{1}{2}\epsilon_{\rho\mu\nu\alpha}u^{\rho}S^{\mu\nu}
\label{pl}
\end{equation}
(In fact, to leading order in the spin $|S|$, the Mathisson-Pirani spin
supplementary condition is identical with the Tulczyjew-Dixon condition
$S^{\alpha\beta}p_{\beta}=0$. That is, the difference between the two is
cubic in the spin of the particle, and hence negligible.) Notice, that the
inverse of Eq.~(\ref{pl}) is given by
\begin{equation}
S^{\alpha\beta}=\epsilon^{\alpha\beta\gamma\delta}u_{\gamma}S_{\delta}
\end{equation}

The Papapetrou equations are then given by
\begin{equation}
\mu\frac{\,Du^{\alpha}}{\,d\tau}=\frac{1}{2}\epsilon^{\lambda\mu\rho\sigma}
R^{\alpha\nu}_{\;\;\;\;\lambda\mu}u_{\nu}u_{\sigma}S_{\rho}
+O(S^2)
\label{papapetrou}
\end{equation}
and
\begin{equation}
\frac{\,DS_{\alpha}}{\,d\tau}=u_{\alpha}S_{\mu}\frac{\,Du^{\mu}}
{\,d\tau}=0+O(S^2)
\label{pt}
\end{equation}  
which means that to leading order in the spin, the Pauli-Lubanski spin
covector is parallel transported along $u^{\alpha}$. (The spin covector is
always Fermi-Walker transported.) In these equations we
neglect all terms at order $S^2$ or higher. From Eq.\
(\ref{papapetrou}) we identify the spin force as
\begin{equation}
f^{\alpha}_{\rm spin}=\frac{1}{2}\epsilon^{\lambda\mu\rho\sigma}
R^{\alpha\nu}_{\;\;\;\;\lambda\mu}u_{\nu}u_{\sigma}S_{\rho}
\end{equation}

\section{Explicit expressions for Schwarzschild}\label{app-sch}
We list in this Appendix the explicit expressions for the solution of the
perturbative equations of motion. We remark that the following explicit
solution appears different from the solutions of Ref.~\cite{burko-03}
because of the different perturbative expansion of the angular velocity
that is used there.
 
\begin{equation} \omega_{(1)}=-\frac{r-3M}{2\mu (Mr)^{1/2}}f^{(1)}_{r}
\end{equation}

\begin{equation}
\omega '_{(1)}=-\frac{1}{4\mu(Mr)^{1/2}}\left[\frac{r+3M}{r}f^{(1)}_{r}
+2(r-3M)f^{(1)'}_{r}\right]
\end{equation}

\begin{equation}
V_{(1)}=\frac{2r}{\mu M}\frac{r-3M}{r-6M}
\left[\left(\frac{M}{r}\right)^{\frac{1}{2}}\left(1-\frac{2M}{r}\right)
f^{(1)}_{\phi}+Mf^{(1)}_{t}\right]
\label{v1-1}
\end{equation}

\begin{eqnarray}\label{v2-1}
V_{(2)}&=&\frac{r(r-3M)}{\mu^2M^2(r-6M)^2}
\left[2\left(\frac{M}{r}\right)^{\frac{1}{2}}f^{(1)}_{\phi}f^{(1)'}_{r}
r(r-2M)^2\right. \nonumber \\
&\times&
(r-3M)+\left(\frac{M}{r}\right)^{\frac{1}{2}}f^{(1)}_{\phi}f^{(1)}_{r}    
(5r-6M)(r-2M) \nonumber \\
&\times& (r-3M)+2Mf^{(1)}_{t}f^{(1)'}_{r}r^2(r-2M)(r-3M) \nonumber \\
&+&4Mf^{(1)}_{t}f^{(1)}_{r}r^2(r-3M) +2\mu M^2f^{(2)}_{t}(r-6M)\nonumber
\\
&+& \left.
2\mu \left(\frac{M}{r}\right)^{\frac{3}{2}}f^{(2)}_{\phi}
(r-2M)(r-6M)\right] 
\end{eqnarray}

\section{The expansion coefficients $C_{k}$}\label{app2}
We fit the luminosity in gravitational waves (data taken from
Ref.~\cite{cutler}) to a function of the form (\ref{fit-pol}) with $n=6$. 
The number of terms we sum over was determined by the accuracy of the
approximation. The coefficients were determined to equal (not all the
figures shown are significant)
\begin{eqnarray}
C_0&=&3.604707682 \nonumber \\ 
C_1&=&-0.9253608775\nonumber \\
C_2&=&0.1300173368\nonumber \\
C_3&=&-9.256723735\times 10^{-3}\\
C_4&=&3.455531402\times 10^{-4}\nonumber \\
C_5&=&-6.363613595\times 10^{-6}\nonumber \\
C_6&=&4.521868774\times 10^{-8}\nonumber \, .
\end{eqnarray}

\end{appendix}

\end{document}